\documentclass[fleqn,usenatbib]{mnras}

\usepackage{newtxtext,newtxmath}
\usepackage{makecell}

\usepackage[T1]{fontenc}

\DeclareRobustCommand{\VAN}[3]{#2}
\let\VANthebibliography\thebibliography
\def\thebibliography{\DeclareRobustCommand{\VAN}[3]{##3}\VANthebibliography}

\usepackage{graphicx}	
\usepackage{amsmath}	

\newcommand{\src}{Swift~J1727}
\newcommand{\lrlx}{$L_R \text{--} L_X$}
\newcommand{\ergs}{erg\,s$^{-1}$}

\title[Swift~J1727.8$-$1613 in the hard state]{The peculiar hard state behaviour of the black hole X-ray binary Swift~J1727.8$-$1613}

\newcommand{\AuthorList}{%
A.~K.~Hughes,$^{1}$\thanks{E-mail: hughesakh@gmail.com}
F.~Carotenuto,$^{2}$
T.~D.~Russell,$^{3}$
A.~J.~Tetarenko,$^{4}$
J.~C.~A.~Miller-Jones,$^{5}$
R.~M.~Plotkin,$^{6,7}$ \newauthor
A.~Bahramian,$^{5}$
J.~S.~Bright,$^{1,8}$
F.~J.~Cowie,$^{1}$
J.~Crook-Mansour,$^{1}$
R.~Fender,$^{1,9}$
J.~K.~Khaulsay,$^{10}$
A.~Kirby,$^{5}$\newauthor
S.~Jones,$^{5}$
M.~McCollough,$^{11}$
R.~Rao,$^{11}$
G.~R.~Sivakoff,$^{12}$
S.~D.~Vrtilek,$^{11}$ 
D.~R.~A.~Williams-Baldwin,$^{13}$\newauthor
C.~M.~Wood,$^{5}$
D.~Altamirano,$^{14}$
P.~Casella,$^{2}$
N.~Castro Segura,$^{15}$ 
S.~Corbel,$^{16}$
M.~Del Santo,$^{3}$ \newauthor
C.~Echibur\'u-Trujillo,$^{17}$
J.~van den Eijnden,$^{15,18}$
E.~Gallo,$^{19}$
P.~Gandhi,$^{14}$ 
K.~I.~I.~Koljonen,$^{20}$  
T.~Maccarone,$^{21}$ \newauthor
S.~B.~Markoff,$^{18,22}$ 
S.~Motta,$^{1,23}$ 
D.~M.~Russell,$^{24}$
P.~Saikia,$^{24}$ 
A.~W.~Shaw,$^{25}$ 
R.~Soria,$^{26,27}$
J.~A.~Tomsick,$^{28}$ \newauthor
W.~Yu,$^{29,30}$
X.~Zhang$^{30}$
}
\newcommand{\EndAffil}{%
$^{1}$Department of Physics, University of Oxford, Denys Wilkinson Building, Keble Road, Oxford OX1 3RH, UK\\
$^{2}$INAF-Osservatorio Astronomico di Roma, Via Frascati 33, I-00078, Monte Porzio Catone (RM), Italy\\
$^{3}$INAF, Istituto di Astrofisica Spaziale e Fisica Cosmica, Via U. La Malfa 153, I-90146 Palermo, Italy\\
$^{4}$Department of Physics and Astronomy, University of Lethbridge, Lethbridge, Alberta, T1K 3M4, Canada\\
$^{5}$International Centre for Radio Astronomy Research, Curtin University, GPO Box U1987, Perth, WA 6845, Australia\\
$^{6}$Department of Physics, University of Nevada, Reno, NV 89557, USA\\
$^{7}$Nevada Center for Astrophysics, University of Nevada, Las Vegas, NV 89154, USA\\
$^{8}$Breakthrough Listen, Astrophysics, Department of Physics, The University of Oxford, Keble Road, Oxford, OX1 3RH, UK\\
$^{9}$Department of Astronomy, University of Cape Town, Private Bag X3, Rondebosch 7701, South Africa\\
$^{10}$Department of Physics and Astronomy, The University of Sheffield, Hicks Building, Hounsfield Road, Sheffield S3 7RH, UK\\
$^{11}$Center for Astrophysics $|$ Harvard \& Smithsonian, 60 Garden Street, Cambridge, Ma, 02138\\
$^{12}$Department of Physics, University of Alberta, CCIS 4-181, Edmonton, AB T6G 2E1\\
$^{13}$Jodrell Bank Centre for Astrophysics, School of Physics and Astronomy, The University of Manchester, Manchester, M13 9PL, UK\\
$^{14}$School of Physics \& Astronomy, University of Southampton, SO17 1BJ, UK\\
$^{15}$Department of Physics, University of Warwick, Gibbet Hill Road, Coventry, CV4 7AL, UK\\
$^{16}$Université Paris Cité and Université Paris Saclay, CEA, CNRS, AIM, F-91190 Gif-sur-Yvette, France\\
$^{17}$Department of Astrophysical and Planetary Sciences, JILA, Duane Physics Bldg., 2000 Colorado Ave., University of Colorado, Boulder, CO 80309, USA\\
$^{18}$Anton Pannekoek Institute for Astronomy, Universiteit van Amsterdam, Science Park 904, 1098, XH, Amsterdam, The Netherlands
$^{19}$Department of Astronomy, University of Michigan, 1085 S University, Ann Arbor, MI 48109, USA\\
$^{20}$Department of Physics, Norwegian University of Science and Technology, NO-7491 Trondheim, Norway\\
$^{21}$Department of Physics and Astronomy, Texas Tech University, Lubbock, TX 79409-1051, USA\\
$^{22}$Gravitation and Astroparticle Physics Amsterdam Institute, University of Amsterdam, Science Park 904, 1098 XH 195 196 Amsterdam, The Netherlands\\
$^{23}$Istituto Nazionale di Astrofisica, Osservatorio Astronomico di Brera, Via E. Bianchi 46, 23807 Merate, (LC), Italy\\
$^{24}$Center for Astrophysics and Space Science (CASS), New York University Abu Dhabi, PO Box 129188, Abu Dhabi, UAE\\
$^{25}$Department of Physics and Astronomy, Butler University, 4600 Sunset Avenue, Indianapolis, IN 46208, USA\\
$^{26}$INAF-Osservatorio Astrofisico di Torino, Strada Osservatorio 20, I-10025 Pino Torinese, Italy\\
$^{27}$Sydney Institute for Astronomy, School of Physics A28, The University of Sydney, Sydney, NSW 2006, Australia\\
$^{28}$Space Sciences Lab, University of California, Berkeley, 7 Gauss Way, Berkeley, CA 94720, USA\\
$^{29}$Shanghai Astronomical Observatory, Chinese Academy of Sciences, Shanghai 200030, China\\
$^{30}$University of Chinese Academy of Sciences, Chinese Academy of Sciences, Beijing 100049, China}

\author[A. K. Hughes et al.]{%
\AuthorList
\\%
Affiliations are listed at the end of the paper
}

\date{Accepted XXX. Received YYY; in original form ZZZ}

\pubyear{\the\year{}}

\begin{document}
\label{firstpage}
\pagerange{\pageref{firstpage}--\pageref{lastpage}}
\maketitle

\begin{abstract}
Tracking the correlation between radio and X-ray luminosities during black hole X-ray binary outbursts is a key diagnostic of the coupling between accretion inflows (traced by X-rays) and relativistic jet outflows (traced by radio). We present the radio--X-ray correlation of the black hole low-mass X-ray binary Swift~J1727.8$-$1613 during its 2023--2024 outburst. Our observations span a broad dynamic range, covering $\sim$4 orders of magnitude in radio luminosity and $\sim$6.5 in X-ray luminosity. This source follows an unusually radio-quiet track, exhibiting significantly lower radio luminosities at a given X-ray luminosity than both the standard (radio-loud) track and most previously known radio-quiet systems. Across most of the considered distance range ($D\,{\sim}\,1.5$--$4.3$\,kpc), Swift~J1727.8$-$1613 appears to be the most radio-quiet black hole binary identified to date. For distances ${\geq}\,4$\,kpc, while \src\ becomes comparable to one other extremely radio-quiet system, its peak X-ray luminosity (${\gtrsim}\,5{\times}10^{38}$\ergs) exceeds that of any previously reported hard-state black hole low-mass X-ray binary, emphasising the extremity of this outburst. Additionally, for the first time in a radio-quiet system, we identify the onset of X-ray spectral softening to coincide with a change in trajectory through the radio--X-ray plane. We assess several proposed explanations for radio-quiet behaviour in black hole systems in light of this dataset. As with other such sources, however, no single mechanism fully accounts for the observed properties, highlighting the importance of regular monitoring and the value of comprehensive (quasi-)simultaneous datasets.
\end{abstract}




\section{Introduction}
A black hole (BH) low-mass X-ray binary (LMXB) is an accreting binary system in which a stellar-mass black hole \citep[typically ${\sim}7\,{\rm M_\odot}$;][]{BlackCAT} accretes material from a low-mass donor star (${\lesssim}1\,{\rm M_\odot}$). As in other accretion-powered systems \citep{livio2002}, the accretion process in BH LMXBs gives rise to highly collimated, relativistic outflows, referred to as \textit{jets}. In these systems, the accretion flow is primarily traced by X-ray emission and the jets are most commonly observed at radio frequencies. Although a robust empirical connection has been established between accretion states and jet activity, the physical mechanisms governing the accretion–ejection coupling remain poorly understood.

BH LMXBs spend the majority of their lifetimes in quiescence at an undetected or low X-ray luminosity \citep[$L_X\,{<}\,10^{34}\,$\ergs;][]{Plotkin2013QS}. As a result, these systems are primarily discovered during their bright active states (`outbursts')  during which their X-ray luminosities can reach values ${\geq}\,10^{38}$\ergs. In outburst, the system evolves through `accretion states', which are defined by the X-ray spectral and timing properties; the two canonical accretion states being the \textit{hard} and \textit{soft} state \citep[see, e.g.,][for reviews]{mclintock2006,Kalemci2022review,DeMarco2022states}. 

In the hard state, the X-ray spectrum is dominated by X-ray photons that have been Compton up-scattered in a coronal flow \citep{Thorne1975,Shapiro1976}, with sub-dominant contributions from an accretion disc \citep{Done2007}. In contrast, the soft state exhibits an inversion of the dominant X-ray component, with the thermalised (multi-colour) accretion disc becoming dominant. An outburst typically begins in the hard state, with rising X-ray luminosity, transitions to the soft state near peak brightness, and returns to the hard state during the decay phase. During transitions, BH LMXBs exhibit intermediate X-ray properties: the hard-intermediate and soft-intermediate states \citep[][]{belloni2016}. While a hard$\rightarrow$soft$\rightarrow$hard evolution is considered typical, some outbursts display repeated transitions through the intermediate state before reaching the soft state, or only remain in a hard state \citep{Brocksopp2001hardonly,btetarenko2016, Alabarta2021a}.

Jets from BH LMXBs are broadly classified as either continuously accelerated, steady `compact jets' or transient `jet ejecta' \citep[see, e.g.,][for reviews]{fender2004,fender2006}. The type of jet produced and the observed radio properties correlate with the accretion state. Compact jets are associated with the hard (and perhaps, hard-intermediate) states. Their emission is consistent with a partially self-absorbed synchrotron spectrum, often exhibiting a flat or mildly inverted spectral index ($\alpha\,{\gtrsim}\,0$; where the flux density, $F_{R,\nu}$, scales with the observing frequency, $\nu$, as $F_{R,\nu} \propto \nu^\alpha$). 

In the soft state, the compact jet emission is quenched \citep[e.g.,][]{Tananbaum1972,Fender1999gx339quench,coriat2011,drussell2011,russell2019,Maccarone2020quench}. However, one or more transient jets may be launched during the hard$\rightarrow$soft transition \citep[e.g.][]{mirabel1994, Fender1999, Brocksopp2002ejection, russel2019, bright2020, Carotenuto2021, wood2021}. Transient jets are discrete, optically thin ($\alpha\,{\lesssim}\,-0.6$) knots of synchrotron-emitting plasma launched from the system at mildly to highly relativistic speeds, decoupled from the accretion flow. 

The radio luminosity of compact jets ($L_R$) exhibits a positive, non-linear correlation with the X-ray luminosity ($L_X$), providing evidence for a coupling between accretion and jet production. This \lrlx\ relation extends across several orders of magnitude in luminosity \citep[e.g.][]{Hannikainen1998b, Corbel2003, gallo2003, gallo2004, corbel2013}, and was initially thought to follow a single power-law relation, $L_X\,{\propto}\, L_R^{\beta}$, with $\beta\,{\sim}\,0.6$---the so-called `standard track' \citep[][]{gallo2003}. With the inclusion of a mass dependence, this relation could be extrapolated to accreting supermassive black holes (i.e., Active Galactic Nuclie; AGN), suggesting a universal scaling \citep[e.g., `the fundamental plane of black hole activity';][]{merloni2003}.

However, it was later discovered that some BH LMXBs display significantly lower radio luminosities at a given $L_X$ \citep[e.g., H1743$-$32;][]{2010jonkerH1743,coriat2011, williams2020}. A broken power law better describes these `outlier' sources---referred to as the `hybrid track'---featuring a steep branch ($\beta\,{\gtrsim}\,1$) at high X-ray luminosities ($L_{X,{\rm tran}}\,{\gtrsim}\,10^{36}$\ergs), which transitions to a shallow branch ($\beta\,{\lesssim}\,0.3$) that rejoins the standard track ($\beta\,{\sim}\,0.6$) at lower luminosities. The X-ray luminosity at which outliers reconnect with the standard track is typically around ${\sim}\,10^{35}\,$\ergs and a radio luminosity of ${\sim}\,10^{29}$\,\ergs\ \citep[e.g.][]{coriat2011, Plotkin2017lrlx}, although some sources rejoin the standard track at significantly lower \citep[e.g.,][]{carotenuto2021b} or higher luminosities \citep[e.g.,][]{Islam2018lrlx, Koljonen2019etemp}. Notably, although termed `outliers', more systems have been observed to follow the hybrid track than the standard track \citep[e.g.,][]{gallo2018}. 

The outliers led to the introduction of `radio-loud' (i.e., standard-track) and `radio-quiet' nomenclature, implicitly treating the standard-track and outlier sources as two distinct populations. It should be noted that `radio-quiet' does not comment on the physical origin of the different populations; it simply means that some sources have lower radio luminosities at a given X-ray luminosity. The reason for the difference may, as some authors have proposed, be due to enhanced X-ray emission from the accretion flow (`X-ray-bright') rather than suppressed radio emission from the jet \citep[e.g.,][]{xie2012lhaf,Meyer-Hofmeister2014,xie2016lhaf}. Moreover, the apparent dichotomy may not reflect two distinct subpopulations \citep{gallo2014, gallo2018}; instead, the allowable \lrlx\ tracks may span the full luminosity range, from the standard track (or above) to extremely radio-quiet behavior, depending on source properties. The physical mechanisms responsible for the radio-loud and radio-quiet sources---whether arising from variations in the jet, the accretion flow, or both \citep[e.g.,][]{coriat2011,Espinasse2018}---remain poorly understood. It is also unclear whether individual sources are intrinsically tied to one track or can transition between radio-loud and radio-quiet behaviour across different outbursts or spectral states, highlighting the need for continued monitoring to disentangle the nature of these ‘populations’.

\subsection{The Black Hole Low-Mass X-ray Binary Swift~J1727.8$-$1613}
Swift~J1727.8$-$1613 (hereafter \src) is a BH LMXB that was discovered in August 2023 during its first recorded outburst. The event lasted approximately one year and was the subject of intensive observational campaigns, particularly at radio and X-ray frequencies \citep[e.g.,][]{Bollemeijer2023a, Bollemeijer2023b, millerjones2023a, millerjones2023b, ingram2023ixpe, Peng2024, Liu2024,Wood2025SWJ1717Ejecta}, making it an excellent candidate to augment our current \lrlx\ catalogues. \src\ swiftly became one of the brightest X-ray sources in the sky \citep[e.g.][]{page2023, Kennea2023}, though it remains unclear whether this brightness is intrinsic or due to its proximity. 

\citet{Mata2024distance1727,Mata2025distance1727} presented optical observations of \src, measuring its radial velocity and dynamically confirming the presence of a black hole ($M_\text{BH}\,{>}\,3\,M_\odot$). Furthermore, these authors estimated a distance of $3.7\,{\pm}\,0.3$\,kpc using an empirical relation between binary period ($P_\text{orb}$ of 10.8\,h) and absolute $r$-band magnitude, together with pre-outburst photometry and optical extinction estimates. \citet{Burridge2025distance1727} reanalysed these data and supplemented them with near-UV spectra, reporting an updated distance of $5.5_{-1.1}^{+1.5}$\,kpc. However, most recently, Burridge et al. (private communication) have revised this estimate to $2.6^{+1.7}_{-1.1}$\,kpc, incorporating updated properties of the companion star. Our analysis adopts the most recent inferred distance range of 1.5–4.3\,kpc, corresponding to the 68$\%$ equal-tailed interval. The estimate reported by \citet{Mata2025distance1727} lies well within this range. We note that distances outside this interval would only make the source appear more extreme, and therefore do not contradict our conclusions.

In this paper, we utilise the comprehensive radio light curves presented in \citet{HughesAKH2025a}, alongside quasi-simultaneous X-ray observations, to investigate the \lrlx\ evolution of \src\ and the connection between its accretion flow and compact jets. In Section~\ref{sec:observations}, we describe the radio and X-ray data selection and analysis. Section~\ref{sec:results} presents the \lrlx\ track, and in Section~\ref{sec:disc_conc} we discuss the physical interpretations.

\section{Observations and Methods}
\label{sec:observations}
This section summarises the observations and analysis used to produce the final data products, all of which---including raw and processed data and analysis scripts---are available on GitHub\footnote{\url{https://github.com/AKHughes1994/SwJ1727_2023_Outburst}}. While \src\ underwent multiple state transitions near the outburst peak \citep[e.g.,][]{Wenfei2023}, we focus only on the hard states immediately following the outburst rise (henceforth `the rise hard state') and during the decay (`the decay hard state') due to the complex radio behavior and jet ejecta contributions during intermediate phases (see Figure~\ref{fig:LC_TOTAL}).

\subsection{Radio Observations}
\label{sec:Radio_OBS}
\src\ was the focus of multiple extensive radio campaigns, which were compiled and presented in \citet{HughesAKH2025a}. In this work, we use a subset of those data, obtained with MeerKAT, the Karl G.~Jansky Very Large Array (VLA), and the Australia Telescope Compact Array (ATCA), as these instruments captured both the rise and decay hard states. 

The \lrlx\ relation pertains specifically to compact jet emission during the hard state. Accordingly, during the rising hard state, we include only those observations taken before the onset of radio flaring (which began on 2023 September 19; MJD\,60206), after which observations with the Very Long Baseline Array revealed that the flares coincided with transient jet ejections \citep[][]{Wood2025SWJ1717Ejecta}. \citet{HughesAKH2025a} presented the entire radio light curve of \src, providing the integrated flux densities, which---during most of the decay hard state---included both unresolved emission from the compact jet (reappearing at the `core' position) and spatially resolved jet ejecta (see Figure~\ref{fig:SWJ1727_image}). The angular extent of (detectable) emission regions from the compact jet is ${\sim}\,0.040^{\prime\prime}$ \citep{Wood2024}, rendering it unresolved at the angular resolution of Meerkat, the VLA, and ATCA (${\gtrsim}\,1^{\prime\prime}$); we measure the compact jet's flux density by fitting only the core emission.

However, throughout the soft state, steep-spectrum (and fading) radio emission was detected at the core position. Since compact jet emission is quenched by $\gtrsim$\,3 orders of magnitude in the soft state (as discussed in Section~1) and, when detected in the hard state, exhibits a flat radio spectrum, this steep-spectrum component most likely originates from long-lived transient jet(s) that remain unresolved from the radio core \citep[similar to those seen in MAXI~J1535$-$571;][]{russel2019}. As a result, following the re-brightening of the compact jet during the decay hard state, the core emission is a superposition of compact jet and ejecta emission. To account for ejecta contributions, we modelled the fading steep-spectrum component with an exponential decay\footnote{A power law model was also fit to the decay; however, given the magnitude of the measurement errors, its fit was indistinguishable from the exponential.}. Extrapolating this decay to the hard state epochs---and accounting for spectral indices---we subtracted the modelled value from the observed core flux density, adopting the resulting value as the true compact jet flux density (see Appendix~\ref{sec:exp_decay_method} for details). Depending on the epoch, the subtracted component constituted 25--40\%, 5--12\%, and 10--37\% of the total for MeerKAT, ATCA, and the VLA, respectively.


We included a systematic error equal to 10$\%$ of the measured flux density (added in quadrature) as fractional errors are often recommended to account for the secular evolution of the calibrator(s) \citep[e.g., 3$\%$ for ATCA;][]{2011MNRAS.415.1597M,2012MNRAS.422.1527M}. While 10$\%$ is overly conservative when considering calibration systematics alone, it should implicitly account for additional systematics from interpolation errors or imperfect subtraction of the decay.

\subsection{X-ray Observations}
\src\ was monitored with the X-ray Telescope onboard the Neil Gehrels \textit{Swift} Observatory \citep[\textit{Swift}-XRT;][]{XRTcite}. We include 20 observations from the rise hard state (2023 Aug 28–Sep 20; MJD 60184–60207, target ID: 1186959\footnote{Labelled as GRB 230824A in the \textit{Swift}-XRT archive due to initial misidentification as a gamma-ray burst.}), and 22 from the decay hard state (2024 Feb 03–Jun 03; MJD 60343–60464, target IDs: 16584, 89766). Spectra were extracted using the \textsc{swifttools} pipeline \citep{Evans2007PipelineA,Evans2009PipelineB} and analysed in \textsc{xspec} \citep{XSPEC}.

Pile-up becomes significant for ${\gtrsim}\,150{\rm\,counts\,s^{-1}}$, and the pipeline includes standard corrections that are generally effective. However, due to its extreme brightness (${\gtrsim}\,5000{\rm\,counts\,s^{-1}}$), the rise hard state data lie in a poorly calibrated regime (P.~Evans, private communication). Consequently, we applied a $10\%$ systematic flux error, added in quadrature with statistical uncertainties; we conservatively apply this to all X-ray epochs, even those unaffected by pile-up.

For spectra with ${\geq}\,500\,$counts, we binned into 25-count intervals (${\geq}$\,20\,bins) and fit using $\chi^2$ statistics with an absorbed power law plus disc blackbody model: \texttt{tbabs} * (\texttt{pegpwrlw} $+$ \texttt{diskbb}), capturing both coronal and disc emission. Absorption was modelled with \texttt{tbabs}, adopting \citet{wilms2000} abundances. After fitting, we measured unabsorbed 1--10 keV flux with the convolution model \texttt{cflux} (i.e., \texttt{tbabs} * \texttt{cflux} * (\texttt{pegpwrlw} $+$ \texttt{diskbb})). We initially let $N_H$ vary, obtaining a variance-weighted average of $(2.68\,{\pm}\,0.02){\times}10^{21}{\rm\,cm^{-2}}$\, consistent with past measurements \citep[e.g.,][]{OconnorSWJ1727Atel,Draghis2023,Peng2024,Stiele2024SwJ1717}, and then fixed $N_H\,{\equiv}\,2.68\times10^{21}{\rm\,cm^{-2}}$ for final fits. For spectra with fewer than 500 counts, we used \texttt{cstat} fitting \citep{CashStats} with single-count binning. In cases where \texttt{cstat} was applied, or when $\chi^2$-binned epochs included fewer than 50 bins, we adopted a simplified power law model: \texttt{tbabs}*\texttt{pegpwrlw}, with fixed $N_H$.

Given the quasi-simultaneity of X-ray and radio data (typically ${\lesssim}\,3\,$days apart), we apply a logarithmic interpolation to map the X-ray fluxes onto the radio epochs. Although approximate, this approach captures the evolution of \src\ (see Figures~\ref{fig:LC_RISE} and \ref{fig:LC_DECAY}).

\subsection{Luminosity Conversion}
As is standard in \lrlx\ analyses \citep[e.g.,][]{arash_bahramian_2022_7059313}, we calculate the 1--10\,keV X-ray luminosities ($L_X$) and 5\,GHz radio luminosities ($L_R$) from the unabsorbed X-ray flux ($F_{\rm X}$) and radio flux density ($F_{R,\nu}$), using $L_X = 4\pi D^2 F_X$ and $L_R = 4\pi D^2 \nu_0 F_{R,\nu_0}$, where $\nu_0 = 5\,$GHz. Since our observations span 0.3--15\,GHz, we scale the flux density measured at the observing frequency $\nu$ using the spectral index $\alpha$: $F_{\nu_0} = F_\nu (\nu_0/\nu)^\alpha$. These luminosities form the basis of the radio--X-ray correlation shown in Figure~\ref{fig:lrlx}.

For epochs with simultaneous multi-frequency coverage, we selected the flux density measurement closest to 5\,GHz. To account for spectral evolution, we calculate inter-band spectral indices from the multi-frequency data. However, most observations during the decay phase were single-frequency. To avoid systematic biases associated with intra-band spectral indices---such as the artificial spectral `flattening' seen in ${<}\,30\sigma$ VLA detections \citep[e.g.][]{Heywood2016}---we interpolate the 1.28\,GHz MeerKAT data to match the ATCA and VLA observation times. This approach allows us to estimate inter-band spectral indices from the interpolated flux densities. For the ATCA observations, which were recorded simultaneously at 5.5 and 9\,GHz, we verified the accuracy of the interpolated spectral index by comparing it with the ATCA-only inter-band spectral index and found good agreement (see Figure~\ref{fig:decay_spectra}, second panel, star markers). We then used the ensemble of inter-band spectral indices to scale all radio flux densities to 5\,GHz, adopting the spectral index measurement closest in time to each corresponding flux density. We include the final \lrlx\ values in Table~\ref{tab:LRLX_table}.

\section{Results}
\label{sec:results}

\begin{figure*}
    \centering
    \includegraphics[width=1.0\linewidth]{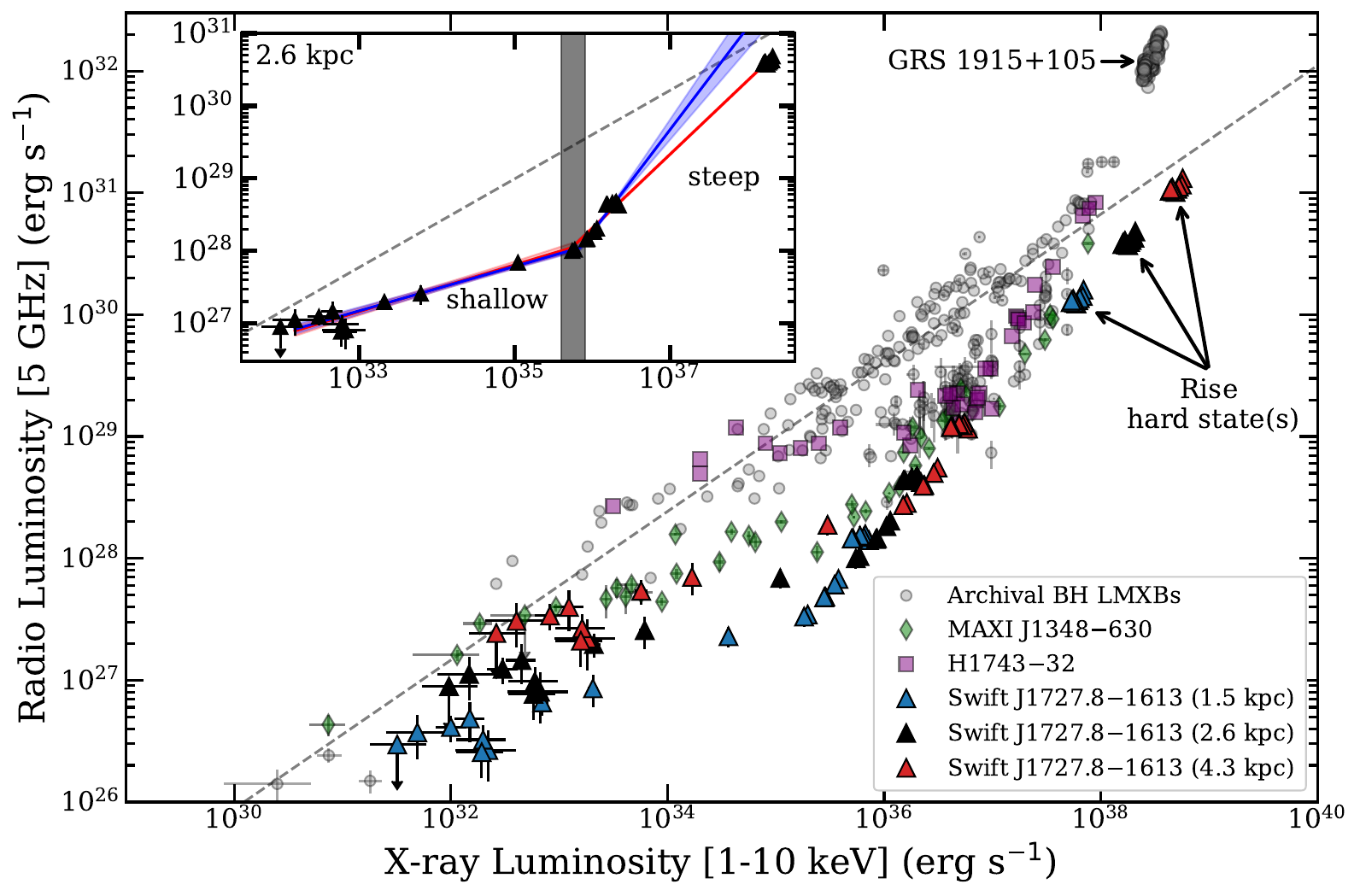}
    \caption{The \lrlx\ evolution of \src\ (triangles) is shown alongside archival BH LMXBs (grey circles; \citealt{arash_bahramian_2022_7059313}), with H1743$-$322 (purple squares) and MAXI~J1348$-$630 (green diamonds) highlighted for comparison. The inset shows a broken power-law fit ($D{=}2.6$\,kpc): red for all data, blue for the decay hard state. The grey shaded region marks the (approximate) 1$\sigma$ range of the transitional luminosity ($L_{X\,\text{tran}}$). The \src\ observations in the upper right-hand corner correspond to the rising hard states. \src’s hybrid track is either an unusually faint decay or an unusually bright peak luminosity in the hard state, depending on the distance.}
    \label{fig:lrlx}
\end{figure*}

In Figure~\ref{fig:lrlx}, we present the \lrlx\ evolution of \src. Due to uncertainties in its distance, we plot three sets of points assuming distances of 1.5, 2.6, and 4.3\,kpc. For context, we include an archival sample \citep[grey circles;][]{arash_bahramian_2022_7059313}, highlighting the archetypal radio-quiet source H1743$-$32 \citep[purple squares;][]{coriat2011}, as well as MAXI~J1348$-$638---which underwent the most radio-quiet evolution of all archival sources to date \citep[green diamonds;][]{carotenuto2021b, Carotenuto2022c}. 

The \lrlx\ evolution reveals that its radio--X-ray correlation cannot be adequately described by a single power law of the form $L_R \,{\propto}\, L_X^\beta$, as is typical for radio-quiet systems. Following the methodology applied to other radio-quiet sources \citep[e.g.][]{coriat2011, carotenuto2021b}, we therefore model the evolution with a broken power law in logarithmic space:
\begin{equation}
    L_R \propto \begin{cases}
    (L_X/L_{X,\text{tran}})^{\beta_\text{shallow}},& \text{if } L_X< L_{X,\text{tran}}\\
    (L_X/L_{X,\text{tran}})^{\beta_\text{steep}},& \text{if } L_X\geq L_{X,\text{tran}}
    \end{cases}.
\end{equation}
We account for uncertainties in both $L_R$ and $L_X$ using orthogonal distance regression, implemented via the \texttt{odr} function from the \textsc{Python} package \textsc{scipy}.

As our monitoring spans both the rise and decay hard states, we fit the \lrlx\ data under two scenarios (adopting $D={2.6}\,$kpc): (i) using both outburst phases, and (ii) using only the decay data, which captures both the steep and shallow tracks (see inset of Figure~\ref{fig:lrlx}). In both cases, the shallow track is robustly constrained and in strong agreement yielding $\beta_\text{shallow} = 0.30\,{\pm}\,0.02$ for (i) and $0.32\,{\pm}\,0.03$ for (ii), which is unsurprising, given that shallow track occurs only during the decaying hard state. The transition luminosities are similarly consistent within uncertainties, with $L_{X,\text{tran}} = (7.1\,{\pm}\,0.9)\times10^{35}$\,\ergs\ for (i), and $(5.4\,{\pm}\,1.1){\times}10^{35}$\,\ergs\ for (ii). The power law indices are independent of the assumed source distance, whereas the luminosities---both at the transition to the shallow track ($L_{X,\text{tran}}$) and when rejoining the standard track ($L_{X,\text{rejoin}}$)---scale with distance as $L_{X,\text{tran}}\,{\approx}\,6{\times}10^{35}(D/2.6{\rm\,kpc})^2$\,\ergs and $L_{X,\text{rejoin}}\,{\approx}\,10^{32}(D/2.6{\rm\,kpc})^2\,$\ergs, respectively. 

As expected, a modest discrepancy (${\lesssim}\,2.7\sigma$) emerges between the steep-track indices, driven by the rising hard state being clustered at luminosities $\gtrsim$\,4 orders of magnitude higher than those observed during the decay. Specifically, we measure $\beta_\text{steep} = 1.02\,{\pm}\,0.01$ for the full dataset and $\beta_\text{steep} = 1.42\,{\pm}\,0.15$ when fitting the decay phase alone. While this difference is $<$\,3$\sigma$, it may hint at mild evolution in the \lrlx\ relation, consistent with prior studies showing limited variability in slope over narrow luminosity ranges \citep{corbel2013}. The tendency for BH LMXBs to appear fainter in radio during the rising hard state may also contribute to the shallower slope when rise data are included \citep{corbel2013,Islam2018lrlx}. Nevertheless, the two fits remain statistically consistent within $2\sigma$, and the derived indices are comparable to those seen in other radio-quiet systems \citep[e.g.,][]{coriat2011,carotenuto2021b}.

\subsection{Distance ambiguity and the extreme nature of Swift~J1727.8$-$1613}
\label{sec:results_extrema}

Regardless of the assumed distance, \src\ exhibits an exceptionally radio-faint \lrlx\ evolution. Initially, \src\ follows the steeper track seen in other radio-quiet sources (e.g., H1743$-$32), before diverging at lower X-ray luminosities and becoming markedly more radio-faint. Near the maximum plotted distance (${\sim}\,4$\,kpc), the source rejoins the standard track at $L_X\,{\lesssim}\,10^{33}$\,\ergs\ ($L_R\,{\sim}\,3{\times}10^{27}$\,\ergs), $\sim$two orders of magnitude below typical rejoin values for radio-quiet systems. At this distance, \src's \lrlx\ behaviour mirrors that of MAXI~J1348$-$630 and, potentially, MAXI~J1631$-$472 if we adopt its minimum distance of 2\,kpc \citep{Monageng2021}. At smaller distances---including the nominal estimate---\src\ becomes the most radio-faint system in the \lrlx\ plane. For instance, at $D\,{\sim}\,1.5$\,kpc, rejoining occurs at $L_X\,{\sim}\,10^{31}$\,\ergs\ ($L_R\,{\sim}\,3{\times}10^{26}$\,\ergs), nearly two orders of magnitude lower in X-rays than even MAXI~J1348$-$630.

The low rejoining luminosity at the nearer distance estimates, together with the apparent consistency with MAXI~J1348$-$630 at larger distances, could be interpreted as supporting a farther distance, i.e., $D\,{>}\,4$\,kpc. However, if the source lies beyond 4\,kpc, its peak X-ray luminosity would be the highest reported for a hard state: $L_{X,\text{max}}\,{\approx}\,5{\times}10^{38}(D/4.3{\rm\,kpc})^2$\,\ergs, surpassing even the exceptionally bright GRS~1915+105 \citep{Rushton2010grs}, although \src\ remains markedly fainter in the radio. Applying a bolometric correction factor of $\tau_\text{bol}\,{\approx}\,2{-}5$ \citep{migliari2006,Anastasopoulou2022bolcorr} to convert from the 1--10\,keV band to the total accretion luminosity, the peak hard state X-ray luminosity is consistent with a $\sim$7\,$M_\odot$ black hole exceeding its Eddington luminosity\footnote{The Eddington luminosity is the luminosity at which radiation pressure overpowers gravity in the limit of a steady, spherical accretion of hydrogen.}: $L_\text{edd} = 1.26{\times}10^{38}\,(M_\text{BH}/M_\odot)$\,\ergs. 

The (bolometric) luminosity at which a state transition occurs is commonly expressed as a fraction of the Eddington luminosity, $f_\text{edd}$. Population studies of BH LMXB outbursts have found the hard$\rightarrow$soft transitions most commonly occur at $f_\text{edd}\,{\sim}\,0.1$ \citep{Dunn2010states,btetarenko2016}. This again suggests that an extreme hard-state luminosity, as implied by a large distance, may be disfavoured. Assuming a typical black hole mass, a closer distance would yield lower X-ray luminosities and a more typical $f_\text{edd}$. However, hard$\rightarrow$soft transition luminosities span a broad distribution, and it is plausible for a system like \src\ to reach Eddington-level luminosities prior to transitioning to the soft state \citep[e.g., a lognormal distribution with $\mu\,{\approx}\,-0.9$, $\sigma\,{\approx}\,0.4$;][]{btetarenko2016}. 

Conversely, the inverse soft$\rightarrow$hard transition tends to occur more narrowly around $f_\text{edd}\,{\sim}\,0.01{-}0.04$ \citep{maccarone2003,Vahdat2019}. For \src, the X-ray luminosity at the time of this transition is ${\approx}\,2{\times}10^{36}(D/2.6{\rm\,kpc})^2$\,\ergs, implying $f_\text{edd}\,{\approx}\,0.007(D/2.6{\rm\,kpc})^2$ (adopting $M_\text{BH}=7\,M_\odot$ and $\tau_\text{bol}=3$). While the nominal distance yields an Eddington fraction consistent with population studies \citep[e.g., the lognormal distribution with $\mu\,{\approx}\,{-}1.8$ and $\sigma\,{\approx}\,0.3$ from][]{Vahdat2019}, smaller distances quickly push the transition luminosity below the observed range for BH LMXBs. This discrepancy is more significant than that introduced by the aforementioned bright hard state, thereby favouring distances at or above the nominal estimate \citep[though we note the existence of outliers with extremely low transitional luminosities, e.g.,][]{Tomsick2014,Chauhan2019}.

In summary, while the distance to \src\ remains uncertain, it is evident that across the range of distances, \src\ stands out as an exceptional system in terms of its \lrlx\ evolution and state transition behaviour.

\subsection{Radio and X-ray spectral evolution}

\begin{figure}
    \centering
    \includegraphics[width=1.0\linewidth]{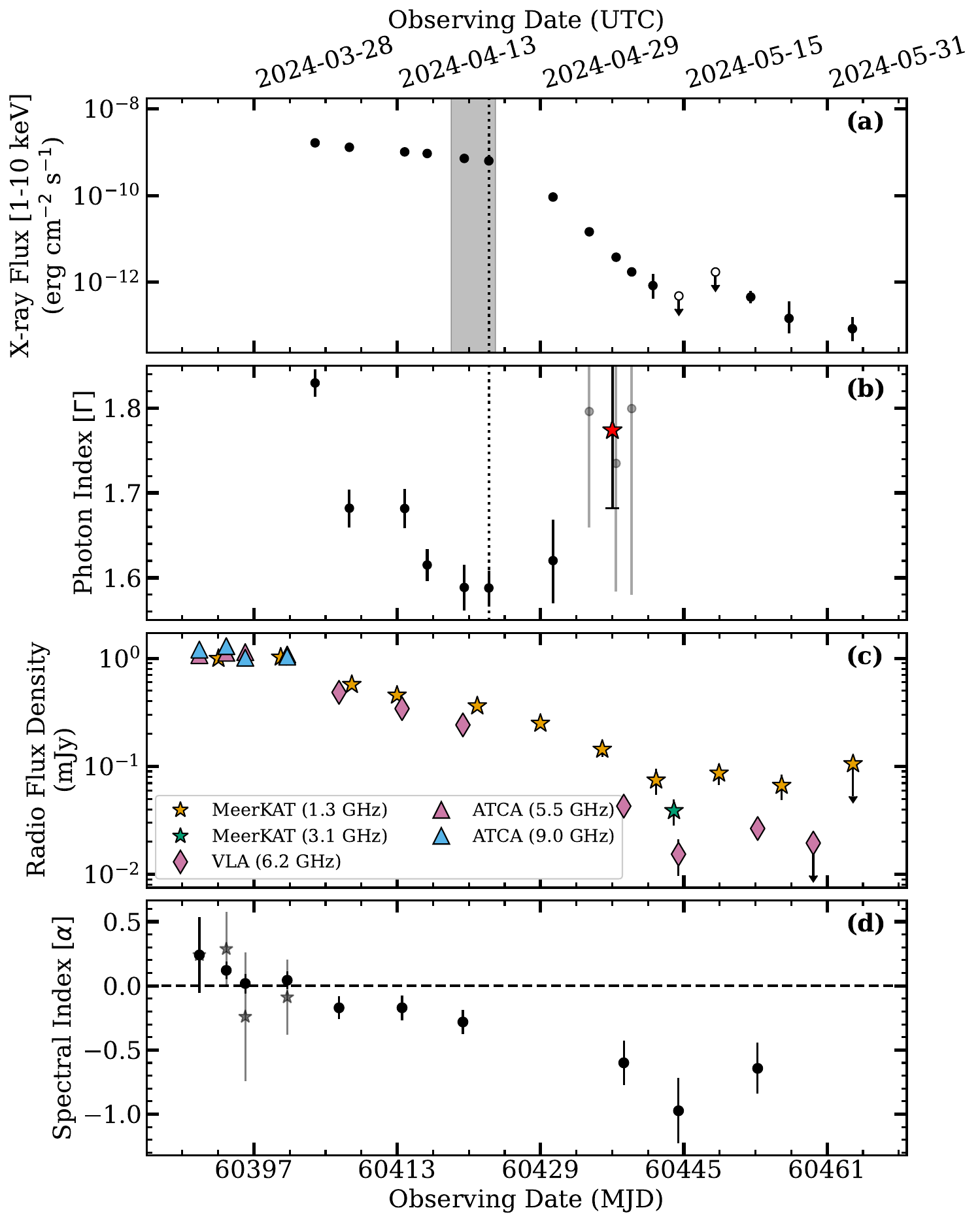}
    \caption{Spectral and photometric evolution of \src\ during the decaying hard state. Panels show: (a) the 1--10\,keV X-ray flux; (b) the X-ray photon index $\Gamma$ (the red star indicates the variance-weighted average of the final three measurements, which had high uncertainties); (c) the radio flux density (marker colour indicates observing frequency; shape indicates facility); and (d) the radio spectral index $\alpha$ (grey stars denote ATCA-only measurements). The vertical dotted line in the top two panels marks the epoch of minimum $\Gamma$. The grey band highlights the times corresponding to the transitional luminosity, as estimated from the decay-only fit, indicating the shift between the steep and shallow tracks. Notable spectral evolution includes the radio spectral index becoming optically thin as the flux density declines, and the source reaching its minimum $\Gamma$ in temporal coincidence with the transitional luminosity.}
    \label{fig:decay_spectra}
\end{figure}

To our knowledge, \src\ is also the radio-quiet system with the best-constrained X-ray spectral evolution during its \lrlx\ evolution.  Like other BH LMXBs (both radio-quiet and radio-loud), the X-ray spectrum hardens to $\Gamma\,{\sim}\,1.5{-}1.6$ during the initial decay, before softening as the source fades into quiescence (top two panels of Figure~\ref{fig:decay_spectra}). Archival BH LMXBs exhibit a plateau at $\Gamma\,{\sim}\,2.1$ as they approach and enter quiescence \citep[e.g.,][]{tomsick01, tomsick04, kong02, corbel06, Plotkin2013QS, Reynolds2014QS}.  Given that our final epoch with spectral constraints yielded $\Gamma\,{\sim}\,1.8$, we expect that \src\ continued to soften after it became too faint for reliable spectral fitting. This places a lower limit of $\gtrsim$\,16\,days on the X-ray softening timescale (measured from the point of minimum $\Gamma$), consistent with other BH LMXBs. However, we note that the few systems with comparable constraints exhibit a broad range of softening timescales \citep[3--90\,days;][]{tomsick01, kalemci05, armaspadilla13, plotkin17V404, beri19, shaw2021}. 

In contrast, a novel result from the \src\ outburst is the temporal coincidence between the minimum $\Gamma$ and the onset of the shallow \lrlx\ track. This alignment is illustrated in Figure~\ref{fig:decay_spectra}a: between 2024 Apr 3--23 (MJD 60403--60423; 20 days), the X-ray flux declined by a factor of $\sim$2 as the spectrum evolved toward its minimum $\Gamma$. Immediately thereafter, over the following $\sim$20 days (2024 Apr 23--May 14; MJD 60423--60444), the X-ray flux dropped by nearly three orders of magnitude, marking the transition onto the shallow track. To highlight this further, we use the interpolated X-ray fluxes to estimate the time range corresponding to the transitional luminosity ($L_{X,\text{trans}}$), based on the decay-only fit; this range is indicated by the vertical grey region in Figure~\ref{fig:decay_spectra}a. A clear temporal alignment is seen between $L_{X,\text{trans}}$ and the epoch of minimum $\Gamma$ (vertical dotted line), reinforcing the link between the spectral minimum and the onset of the shallow track. This suggests that, in the case of \src, the steep-to-shallow transition marks a genuine physical change in the accretion flow, consistent with multi-outburst analyses of H1743$-$32 \citep{Cao2014}, but here observed within a single outburst. Other sources exhibiting spectral evolution during outburst decay have not shown such transitions. For instance, MAXI J1820$+$070 remained on the standard track throughout its evolution \citep{shaw2021}.

Accompanying the X-ray spectral evolution during the decay, the radio spectral index evolved from $\alpha\,{\sim}\,0$ (self-absorbed) to ${\sim}\,{-}0.7$ (optically thin; bottom two panels of Figure~\ref{fig:decay_spectra}). This change may reflect intrinsic variations in the compact jet \citep[e.g.,][]{Corbel2013CJFormation,Plotkin2019spix, russel2020}. Alternatively, the compact jet may have faded, and we are now sampling residual (unsubtracted) ejecta emission. The latter scenario implies an even more radio-faint jet; however, our data do not allow us to distinguish between these possibilities.

\section{Discussion}
\label{sec:disc_conc}

We have presented the \lrlx\ evolution of \src\ during its 2023--2024 outburst. Our coverage spans $\sim$6.5 orders of magnitude in X-ray luminosity and $\sim$4 in radio luminosity, revealing either an exceptionally X-ray bright rising hard state or the most radio-quiet decays observed, depending on the distance.

While the outlier track is considered standard \lrlx\ behaviour for radio-quiet sources, to date, there are only three (four with the inclusion of \src) BH LMXBs with monitoring that captured both the steep and shallow branches \citep[H1743$-$32, Swift J1735.5$-$0127, and MAXI J1348$-$630;][]{coriat2011,Plotkin2017lrlx,carotenuto2021b}. There is no consensus on the mechanisms that cause the radio-quiet and radio-loud evolutions, or whether they arise from differences in the properties of the jets or the accretion flows (or a combination of multiple factors). Here, we briefly discuss some of the explanations reported in the literature, considering their feasibility in light of the results presented here.  

\subsection{Jet-based origins}
Several mechanisms have been proposed to explain the observed radio-quiet and radio-loud populations. Some attribute these differences to variations in the jet properties, which may be influenced by both intrinsic factors and observational biases. \citet{Espinasse2018} investigated the correlation between a system’s radio spectral index and its classification as radio-loud or radio-quiet. The authors found statistically significant differences between the two populations: radio-quiet sources tended to have steeper spectral indices on average ($\alpha_\text{avg}\,{\sim}\,{-}0.3$) compared to the radio-loud population ($\alpha_\text{avg}\,{\sim}\,0.2$). Applying this interpretation to \src\ is not straightforward, as its spectral index varied significantly during the decay ($-1\,{\lesssim}\,\alpha\,{\lesssim}\,0.3$), reaching optically thin values at later epochs. As discussed in Section~3.2, we cannot rule out the possibility that the late-time optically thin spectral index resulted from unsubtracted ejecta emission. However, given that \src\ exhibited a flat spectral index during the early stages of the decay, if the steepening is indeed due to ejecta, this may imply that the compact jet remained self-absorbed (and spectrally flat) throughout, making \src\ an outlier with respect to the findings of \citep{Espinasse2018}.

Alternatively, \citet{Motta2018} proposed that the standard and hybrid \lrlx\ tracks could result from differences in viewing angle, rather than intrinsic differences between individual BH LMXBs. Their analysis revealed that radio-loud systems tend to have lower inclination angles than their radio-quiet counterparts, enhancing radio luminosity due to relativistic beaming. In the case of \src, modelling of its X-ray spectra \citep[e.g.,][]{Peng2024,Svoboda2024,Liu2024} suggests a moderate inclination ${\sim}\,30{-}50^{\circ}$---comparable to estimates for most radio-loud sources \citep[e.g.][]{Motta2018}. However, inclinations derived from X-ray spectral modelling are highly model-dependent. A more robust, albeit less constraining, upper limit of ${<}\,74^{\circ}$ was inferred from the absence of optical variability due to gravitationally driven warping of the companion star \citep[i.e., ellipsoidal modulation;][]{Mata2025distance1727}.

In extreme systems, outbursts of low-inclination, highly beamed BH LMXBs have shown enhanced radio luminosities relative to the standard \lrlx\ track, particularly at higher X-ray luminosities. This deviation is typically attributed to an increasing jet speed---and thus increased beaming---with rising accretion rates \citep[e.g.,][]{Russell2015maxij1836LrLx,Zhang2025lrlx}. However, even for distances corresponding to extreme X-ray luminosities, no radio enhancement was observed for \src; its rise hard state remains consistent with, or even below, the standard track. Assuming that relativistic beaming should affect the hard-state radio luminosities (i.e., its inclination is not near ${\sim}\,70^\circ$), this could imply that \src's compact jet is either intrinsically less relativistic than those of other low-to-moderate inclination systems, or that other factors more strongly modulate the observed radio emission \citep[e.g. the stratified absorption scenario proposed by][]{Motta2018}. Given the totality of evidence, inclination alone does not appear to account for the radio-quiet nature of \src.

Magnetic fields are another property that could account for radio-quiet behaviour. \citet{casella2009} \citep[and][]{2009Peer} proposed that strong internal magnetic fields within compact jets could suppress the observable radio emission. Unfortunately, we cannot assess the viability of this scenario from the \lrlx\ evolution alone. A more comprehensive investigation of the entire outburst---particularly the polarimetric evolution, which directly probes the magnetic field---may be essential to test this hypothesis. Strong magnetic fields remain a plausible explanation for the observed `radio-quietness' of \src\ and potentially other BH LMXBs.

\subsection{Accretion-based origins}
\label{sec:xray_loud}
It is also possible that the apparent `radio-quietness' arises not from a suppression of jet power, but rather from an enhancement in accretion luminosity. Radiatively efficient flows --- such as luminous hot accretion flows (LHAFs; \citealt{yuan2001lhaf}) convert a larger fraction of the rest mass energy into radiation than radiatively inefficient flows, which predominantly advect energy across the event horizon or expel it via outflows (e.g., advection-dominated accretion flows, ADAFs; \citealt{Narayan1994adaf}). The radiative efficiency can be parameterised by $\eta$ where $L_X = \eta \dot{M} c^2$, and $\dot{M}$ is the mass accretion rate. In LHAFs, the radiative efficiency $\eta$ can exceed that of ADAFs by an order of magnitude \citep{xie2012lhaf, xie2016lhaf}. These more efficient accretion flows have also been proposed to explain the distinct radio--X-ray correlations observed in AGN \citep{Dong2014AGNefficient, Panessa2015AGNefficient}. However, caution is warranted when comparing black hole LMXBs to AGN, as the latter---depending on their subclass---may exhibit significantly different accretion geometries and therefore may not serve as direct analogs to hard-state BH LMXBs (e.g., LHAFs versus thin disks).

Within this efficiency framework, there exists a critical mass accretion rate ($\dot{M}_C$) at which a system undergoes a transition to higher radiative efficiency \citep{Narayan1998LC,yuan2001lhaf,xie2012lhaf,xie2016lhaf}: $\dot{M}_C\,{\approx}\,5\alpha_v^2\theta_e^{3/2}\dot{M}_\text{edd}$, where $\alpha_v\,{\sim}\,0.1{-}1$ \citep[e.g.,][]{btetarenko2018winds} characterizes the efficiency of angular momentum transport, $\theta_e\equiv k_BT_e/m_ec^2$ is the dimensionless electron temperature \citep[e.g., $kT_e\,{\sim}\,10{-}1000\,$keV;][]{Koljonen2019etemp,Yan2020etemp}, and $\dot{M}_\text{edd} = 10L_\text{edd}/c^2$ is the Eddington mass accretion rate. 

As $\dot{M}{\rightarrow }\dot{M}_C$, the radiative efficiency increases rapidly (e.g., ${<}1\%$ to ${\sim}\,8\%$; \citealt{xie2012lhaf}). Once $\dot{M}\,{>}\,\dot{M}_C$, the flow becomes radiatively efficient, steepening the radio/X-ray correlation, $L_R\,{\propto}\,L_X^{{\sim}1.3}$ \citep[see, ][and references therein]{coriat2011}. In this picture, the shallow track represents an intermediate regime, where the system transitions between the standard (radiatively inefficient) and steep (radiatively efficient) tracks. During this phase, $L_X$ increases more rapidly than $L_R$ for a given rise in $\dot{M}$ \citep{xie2016lhaf}. Systems with larger $\alpha_v$ or $\theta_e$ values exhibit higher critical accretion rate thresholds, $\dot{M}_C$, and may remain in a radiatively inefficient state for most---or even all---of an outburst, thereby appearing radio-loud. Recent broadband X-ray analyses have found evidence for a steeper \lrlx\ track at high X-ray luminosities (${\gtrsim}\,5{\times}10^{37}$\,\ergs) in the canonical radio-loud source GX~339$-$4, supporting the interpretation that radio-loud evolutions can result from systems with significantly larger $\dot{M}_C$ values \citep{Islam2018lrlx,Koljonen2019etemp}. 

\subsubsection{Did Swift~J1727.8$-$1613 contain a radiatively efficient flow?}

Intriguingly, the minimum observed $\Gamma$ coincides with the inflection point in \lrlx\, where the correlation transitions from $\beta_\text{steep}$ to $\beta_\text{shallow}$. For the first time in a radio-quiet system, we find evidence for the shallow \lrlx\ track coinciding with the onset of X-ray spectral softening. While such softening could arise from either inefficient accretion or jet-related processes \citep[see, e.g.][]{corbel06, Sobolewska2011, Plotkin2013QS}, the alignment of the spectral transition and the `track switch' strongly favours an accretion-driven origin. In particular, a jet-based explanation---such as the onset of synchrotron cooling at high energies---would predict a steeper correlation slope of $\beta\,{>}\,1$ in \lrlx\ \citep{yuan05}, contrary to what is observed. Thus, in \src, the X-ray softening appears consistent with the accretion flow evolution driving the shallow \lrlx\ track---possibly linked to the radiative efficiency transition discussed above.

Recognizing that the efficiency transition occurs during the shallow branch (i.e., when $L_X\,{\lesssim}\,L_{X,\text{tran}}$) we can use the relation $L_X = \eta\dot{M}c^2$ and the definition of $\dot{M}_C$, to derive the following expression:
\begin{align}
    M_\text{BH}\,{\lesssim}\,10^{-4}\eta^{-1}\alpha_v^{-2}\theta_e^{-3/2}\left(\frac{D}{2.6{\rm\,kpc}}\right)^2\,M_\odot.
\end{align}
Although $M_\text{BH}$, $\alpha_v$, $\theta_e$, and $D$ are not currently known for \src---nor is any potential temporal evolution of these properties---independent measurements of any subset can help constrain the others. For example, adopting $\alpha_v\,{\gtrsim}\,0.6$ and $kT_e\,{\gtrsim}\,60$\,keV at a lower-bound distance of $D\,{\simeq}\,1.5$\,kpc predicts $M_\text{BH}\,{\lesssim}\,1\,M_\odot$ (assuming $\eta = 0.01$), which is inconsistent with even the lowest plausible mass for a stellar-mass black hole. Other combinations yield more realistic limits; for instance, $\alpha_v\,{\sim}\,0.3$ and $kT_e\,{\sim}\,50$\,keV at $D\,{\simeq}\,2.6$\,kpc predict $M_\text{BH}\,{\lesssim}\,20\,M_\odot$. Multiwavelength constraints on key parameters---particularly distance---are therefore essential to determine the nature of this system. More broadly, such constraints across a larger sample are critical for understanding the physical origins of the dramatic range in transitional luminosities \citep[i.e. spanning more than six orders of magnitude;][]{Islam2018lrlx, Koljonen2019etemp, carotenuto2021b}.

At all considered distances, the inferred (1--10\,keV) transitional luminosity exceeds $5{\times}10^{35}$\,\ergs, corresponding to $f_\text{edd}\,{>}\,0.008$ (assuming $\tau_\text{bol} = 3$ and $M_\text{BH} = 7\,M_\odot$). At these accretion rates, numerical simulations show that radiatively inefficient flows become unsustainable, indicating that \src\ likely hosts a radiatively efficient accretion mode (at some point) during the outburst. However, this raises a deeper question: how do some systems remain on the standard (radio-loud) track well into the hard state, even at luminosities far exceeding this threshold (e.g., for $f_\text{edd}\,{>}\,0.1$)? Indeed, current models struggle to reproduce the behavior of X-ray binaries in this regime \citep[see][and references therein]{Skadowksi2017}.

It should be noted that our discussion is not intended as an exhaustive summary of the jet- or accretion-based mechanisms capable of producing the variation in \lrlx\ tracks, and the scenarios considered here---and those in the broader literature \citep[e.g.,][]{Meyer-Hofmeister2014}---should not be viewed as mutually exclusive. Given the complexities of BH LMXBs, it is likely that multiple processes act in concert to enforce consistency with, or drive deviations from, the `standard' \lrlx\ track, and their relative influence may vary throughout the outburst. While the precise mechanism(s) responsible for the behaviour of \src\ remain uncertain, comprehensive datasets tracking anomalous behaviours such as this provide vital opportunities for testing our models and advancing our understanding of jet–accretion coupling.

\section{Conclusion}

We have presented the radio--X-ray evolution of the black hole low-mass X-ray binary Swift J1727.8$-$1613 during its 2023--2024 outburst. Our analysis, based on an extensive set of quasi-simultaneous radio and X-ray observations, allowed us to track the jet--accretion coupling over $\sim$6.5 orders of magnitude in X-ray luminosity and $\sim$4 in radio luminosity.

\begin{itemize}
    \item The source exhibited a broken power-law evolution with well-constrained steep and shallow branches, characteristic of a hybrid (radio-quiet) track.
    \item For the first time in a radio-quiet system, we found the transition to the shallow track coincided with the onset of X-ray spectral softening, suggesting a link to changes in the accretion flow.
    \item The radio–X-ray track of Swift~J1727.8$-$1613 is highly sensitive to the assumed distance. At $D\,{\sim}\,1.5$--$4.0$\,kpc, the source re-joins the standard track at extremely low luminosities ($L_X\,{\sim}\,10^{31}$--$10^{32}$\,erg\,s$^{-1}$), making it the most radio-quiet (BH LMXB) evolution identified to date. However, at the upper end of the credible distance range ($D \gtrsim 4.0$\,kpc), its radio luminosity becomes broadly consistent with sources like MAXI~J1348$-$630, placing Swift~J1727 near the faint edge of the archival radio-quiet population.
    \item If Swift~J1727.8$-$1613 lies near at or beyond the upper end of the distance range ($D\,{\gtrsim}\,4$\,kpc), its peak 1--10\,keV X-ray luminosity reaches $L_X\,{\gtrsim}\,5{\times}10^{38}$\ergs. Applying a typical bolometric correction factor ($\tau_\mathrm{bol}\,{\sim}\,2$--$5$), this implies a bolometric luminosity exceeding the Eddington luminosity of a ${\sim}\,7\,M_\odot$ black hole. In this scenario, \src\ would represent the most X-ray luminous hard-state BH LMXB observed to date in the \lrlx\ plane.
\end{itemize}

While multiple jet- and accretion-based mechanisms have been proposed to explain radio-quiet behaviour, it remains unclear which (or what combination of) mechanisms can fully account for \src's properties. The coincidence of X-ray softening with the transition to the shallow \lrlx\ track favours, in part, an accretion-driven evolution---possibly tied to a change in radiative efficiency---over purely jet-based explanations. Future constraints on the distance, multiwavelength spectral modelling, and polarimetric monitoring will be key to clarifying the physical drivers behind the diversity in jet--accretion coupling observed across BH LMXBs.

\section*{Acknowledgements}

AKH thanks UKRI for support. AJT acknowledges that this research was undertaken thanks to funding from the Canada Research Chairs Program and the support of the Natural Sciences and Engineering Research Council of Canada (NSERC; funding reference number RGPIN--2024--04458). FC acknowledges support from the Royal Society through the Newton International Fellowship programme (NIF/R1/211296). TDR is an INAF IAF fellow. RS acknowledges the INAF grant number 1.05.23.04.04. JvdE acknowledges a Warwick Astrophysics prize post-doctoral fellowship made possible thanks to a generous philanthropic donation, and was supported by funding from the European Union's Horizon Europe research and innovation programme under the Marie Sklodowska-Curie grant agreement No 101148693 (MeerSHOCKS) for part of this work. CMW acknowledges financial support from the Forrest Research Foundation Scholarship, the Jean-Pierre Macquart Scholarship, and the Australian Government Research Training Program Scholarship. This project has received funding from the European Research Council (ERC) under the European Union’s Horizon 2020 research and innovation programme (grant agreement No. 101002352, PI: M. Linares). RF thanks UKRI, The ERC and The Hintze Family Charitable Foundation for support. RMP acknowledges support from NASA under award No. 80NSSC23M0104. DMR and PS are supported by Tamkeen under the NYU Abu Dhabi Research Institute grant CASS. GRS is supported by NSERC Discovery Grant RGPIN-2021-0400.

The National Radio Astronomy Observatory is a facility of the National Science Foundation operated under cooperative agreement by Associated Universities, Inc. The MeerKAT telescope is operated by the South African Radio Astronomy Observatory, which is a facility of the National Research Foundation, an agency of the Department of Science and Innovation. We acknowledge the use of the Inter-University Institute for Data Intensive Astronomy (IDIA) data intensive research cloud for data processing. IDIA is a South African university partnership involving the University of Cape Town, the University of Pretoria and the University of the Western Cape. The Australia Telescope Compact Array is part of the Australia Telescope National Facility which is funded by the Australian Government for operation as a National Facility managed by CSIRO. We acknowledge the Gomeroi people as the traditional owners of the ATCA observatory site.

\section*{Data Availability}
Data from the VLA are available through the VLA data archive (Project IDs: 23A–260, 2B–069, and 3B–064): \url{https://data.nrao.edu/portal/}. Data from MeerKAT are available through the SARAO data archive (Proposal IDs: 
SCI-20180516-PW-01 and SCI-20230907-RF-01): \url{https://archive.sarao.ac.za/}. Data from ATCA are available through the ATNF archive (Project codes: C2601, C3057, and C3362): \url{https://atoa.atnf.csiro.au}. Data from the \textit{Swift}-XRT are publicly available through the \textit{Swift} archive: \url{https://www.swift.ac.uk/swift_portal}. We host machine-readable data files and analysis scripts at \url{https://github.com/AKHughes1994/SwJ1727_2023_Outburst}.



\bibliographystyle{mnras}
\bibliography{bibly} 

\ \par
{\itshape
\noindent\EndAffil}



\appendix

\section{Light Curves and Interpolation}

\begin{figure*}
    \centering
    \includegraphics[width=1.0\linewidth]{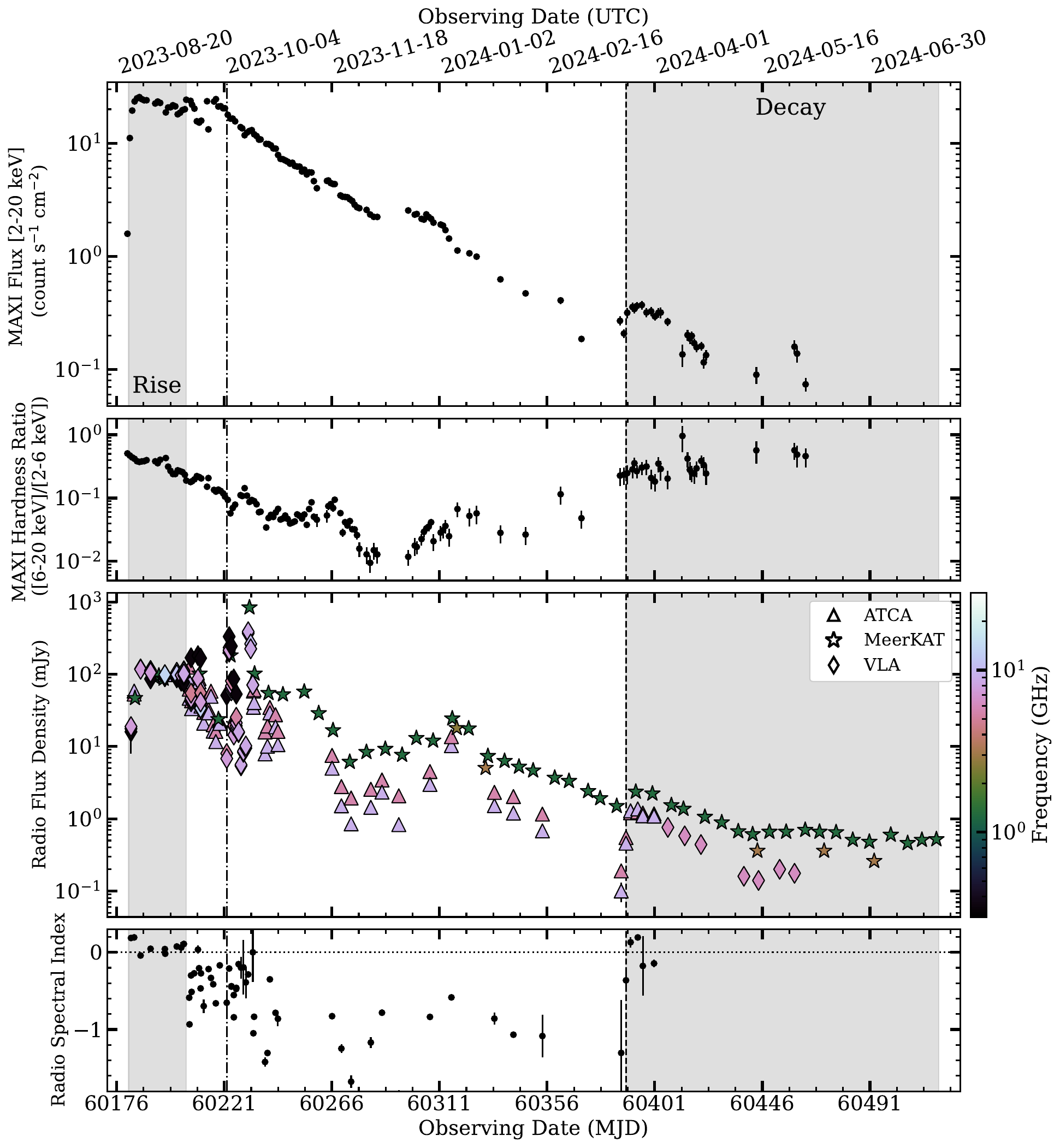}
    \caption{Radio and X-ray light curves covering the entirety of the 2023--2024 outburst of \src: MAXI/GSC 2--20\,keV X-ray flux (\textit{top panel}); X-ray hardness ratio calculated from the 2--6\,keV and 6--20\,keV sub-bands (\textit{second panel}); MeerKAT, VLA, and ATCA light (multi-frequency) light curves (\textit{third panel}); inter-band radio spectral indicies (\textit{bottom panel)}. The dashed-dotted and dashed lines correspond to the hard$\rightarrow$soft and soft$\rightarrow$hard transitions reported in the literature \citep{Bollemeijer2023a,Bollemeijer2023b,Podgorny2024,Russel2024}. The grey shaded regions highlight the times used for the rising and decaying hard states presented in this work. We truncated the rise hard state before the reported state transition due to the onset of flaring and transient jet ejections \citep{Wood2025SWJ1717Ejecta}.}
    \label{fig:LC_TOTAL}
\end{figure*}

Figure~\ref{fig:LC_TOTAL} presents a multi-wavelength photometric and spectral view of the 2023--2024 outburst of \src. From top to bottom: (i) the 2--20\,keV flux from the Monitor of All sky X-ray Image Gas Slit Camera \citep[MAXI/GSC, \textit{top panel;}][]{MAXI2009,maxigsc}; (ii) the X-ray hardness ratio, defined as the ratio of flux from the 2--6\,keV and 6--20\,keV MAXI/GSC fluxes (\textit{second panel)}; (iii) the Meerkat, VLA, and ATCA light curves; and (iv) the inter-band radio spectral indices from simultaneous multi-frequency VLA and ATCA observations (\textit{bottom panel}).

Shaded grey indicates the rise and decay hard states. The vertical dot-dashed and dashed lines mark the reported hard$\rightarrow$soft and soft$\rightarrow$hard transitions, respectively \citep{Bollemeijer2023a,Bollemeijer2023b,Podgorny2024,Russel2024}. As discussed in the main text, we truncate the rise hard state earlier than the reported transition to avoid contamination from flaring activity that launches multiple jet ejecta \citep{Wood2025SWJ1717Ejecta}. These ejections, unresolved at the angular resolution of our instruments, would be spatially coincident with the compact jet. Their presence is reflected in the radio spectral index evolution (\textit{bottom panel}), which displays significant variability between the rise hard state grey region and the reported transition, indicative of a rapidly evolving population of jet components. 

\begin{figure}
    \centering
    \includegraphics[width=1.0\linewidth]{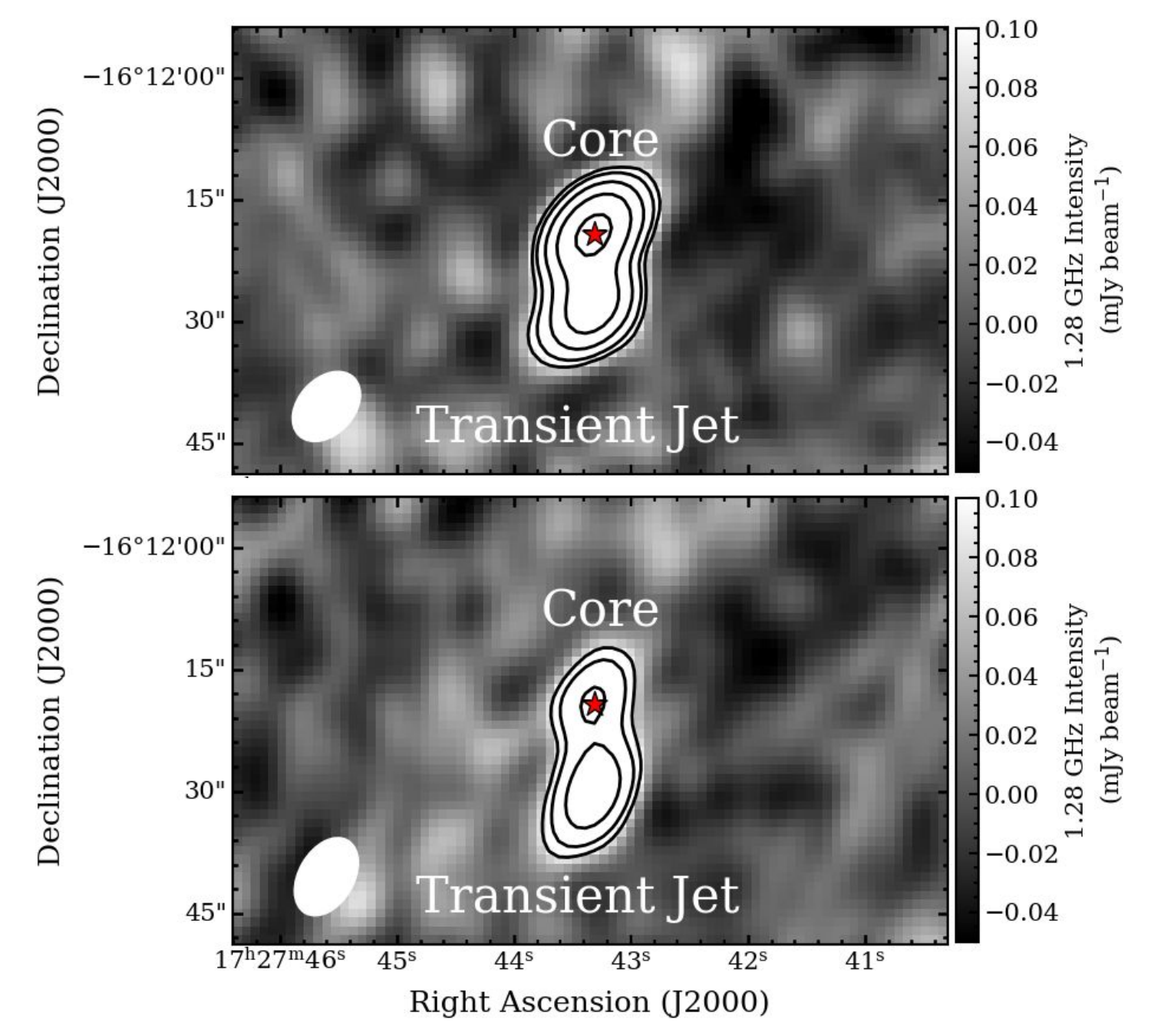}
    \caption{Sample radio images of \src\ from: 2024 March 4 (MJD 60373) while the source was in the soft state (\textit{top panel}); and 2024 April 28 (MJD 60428) after the soft$\rightarrow$hard transition (i.e., the decay hard state, \textit{bottom panel}). The red star indicates the radio position of the core, as measured by the VLA and reported in \citet{millerjones2023a}. Even while \src\ was in the soft state, during which compact jet emission is quenched, there is still bright core emission from unresolved jet ejecta. Additionally, there is a resolved transient jet south of the core position. This resolved jet dominates the integrated flux density at the end of the outburst.}
    \label{fig:SWJ1727_image}
\end{figure}

\begin{figure}
    \centering
    \includegraphics[width=1.0\linewidth]{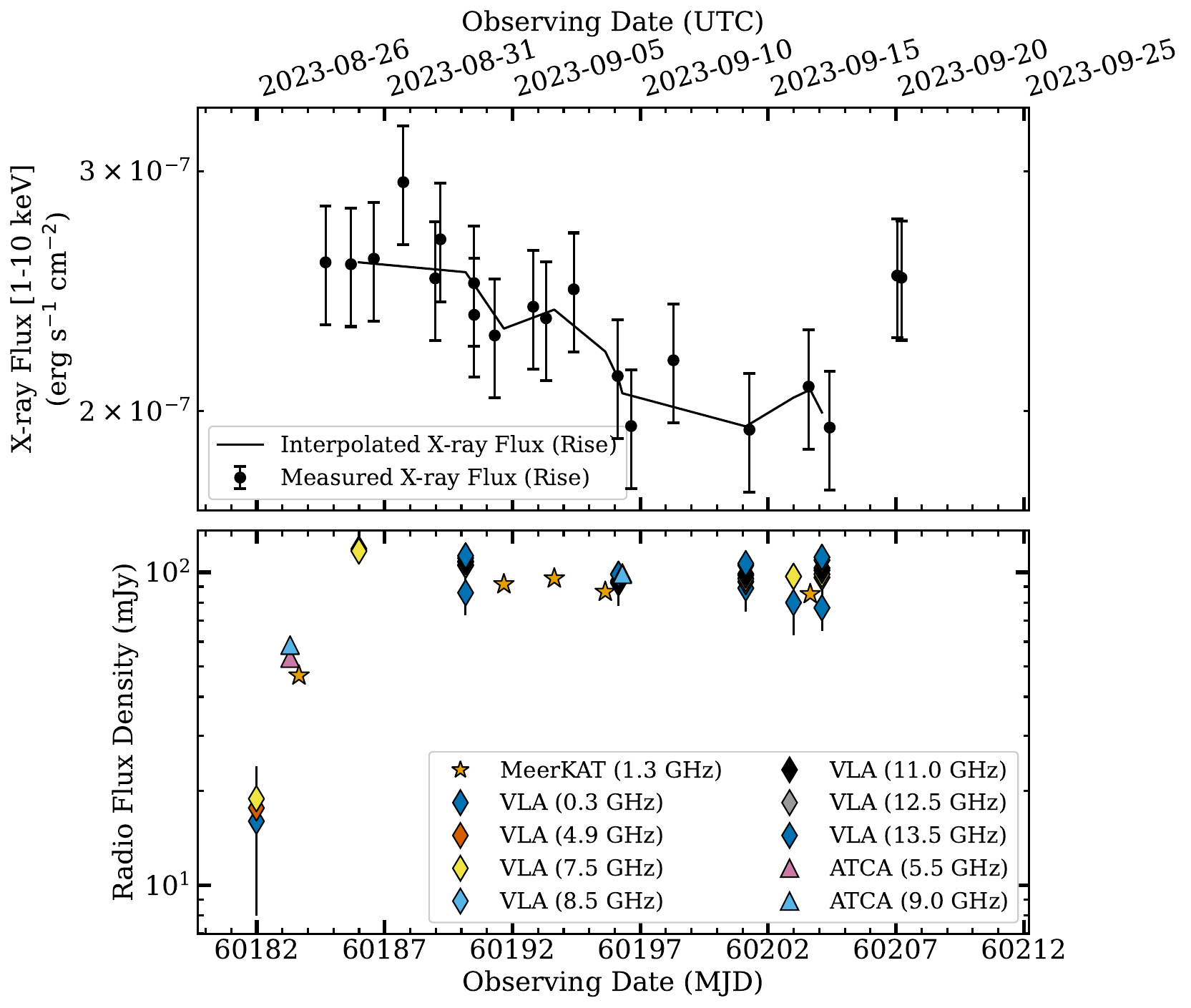}
    \caption{Light curves showing the (\textit{Swift}-XRT) 1--10\,keV X-ray flux (\textit{top panel}) and core radio flux densities (\textit{bottom panel}) during the rise hard state of \src. Solid lines in the top panel trace X-ray fluxes interpolated onto the radio observation epochs (with data points marked by vertices), while black circles indicate the (unabsorbed) X-ray fluxes as measured by \textit{Swift}-XRT. Interpolation was performed in log-space. For clarity, error bars on the interpolated X-ray fluxes are omitted, but those on the measurements are representative of the adopted uncertainties.}
    \label{fig:LC_RISE}
\end{figure}

\begin{figure}
    \centering
    \includegraphics[width=1.0\linewidth]{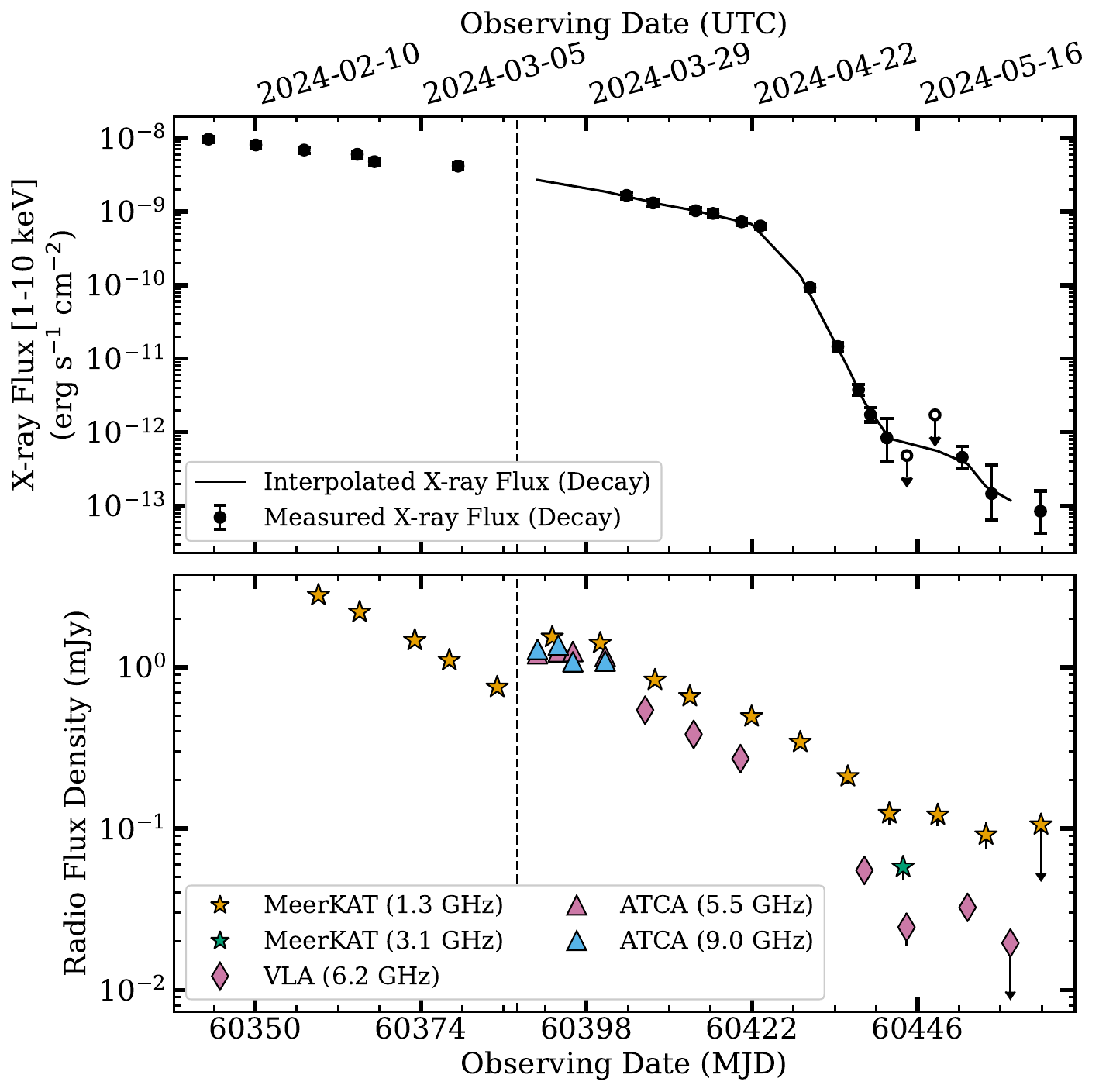}
    \caption{Same as Figure~\ref{fig:LC_RISE} but for the decay hard state. The vertical dashed line market the soft$\rightarrow$hard state transition \citep{Podgorny2024,Russel2024}.}
    \label{fig:LC_DECAY}
\end{figure}

The radio light curves shown in \citet{HughesAKH2025a} represent the total integrated flux densities. Consequently, for epochs with multiple spatial components (e.g., the compact jet and discrete ejecta), the fluxes in Figure~\ref{fig:LC_TOTAL} reflect the sum of multiple components. While the (included) rising hard state only includes compact jet emission, the decay hard state features bright, resolved southern-moving ejecta (Figure~\ref{fig:SWJ1727_image}), which become the dominant radio source at late times. This leads to the plateau seen in the radio decay phase of Figure~\ref{fig:LC_TOTAL}. For a clearer view of the compact core evolution, we present core-only light curves in Figures~\ref{fig:LC_RISE} and \ref{fig:LC_DECAY}. Furthermore, we show both the measured and interpolated X-ray flux values, confirming that our linear interpolation method neither introduces nor smooths over real X-ray variability.

\section{Exponential Fit and Decay Subtraction}
\label{sec:exp_decay_method}
The image morphology (Figure~\ref{fig:SWJ1727_image}) and radio light curves (Figure~\ref{fig:LC_DECAY}) indicate the presence of core emission during the soft state. Since compact jets are expected to be quenched in this state, and the observed core emission is spectrally steep ($\alpha\,{\sim}\,-1$, third panel of Figure~\ref{fig:LC_TOTAL}), we interpret the soft state core emission as arising from unresolved jet ejecta. After compact jet reformation, the core emission becomes a superposition of components. However, only the compact jet contribution is relevant for the \lrlx\ plane. 

To isolate the compact jet emission, we modelled the fading ejecta as an exponential decay, 
\begin{align}
    F_{R,\nu_0} = Ae^{-\frac{t}{\tau}},
\end{align}
where $\nu_0 = 1.28$\,GHz corresponds to the MeerKAT observing frequency. We fit this model using the final five Meerkat epochs and subtracted the resulting flux from each observation to estimate the compact jet contribution.  We find a decay timescale of $\tau = 20.5\,{\pm}\,0.8$,days, similar to previous measurements of BH LMXB ejecta decays \citep[e.g. $\tau = 21.0\,{\pm}\,0.9$,days; see extended data Fig.3 of][]{bright2020}. Because our radio observations span a wide range of frequencies ($\nu\,{\sim}\,1{-}10$\,GHz), we scaled the model flux to the appropriate observing frequency using a power law spectrum: $F_{R,\nu} = F_{R,\nu_0}(\nu/\nu_0)^{\alpha_\text{ej}}$. We adopt $\alpha_\text{ej} = -1$, consistent with the inter-band ATCA spectral indices measured at the end of the soft state (third panel of Fig.~\ref{fig:LC_TOTAL}).

\begin{figure}
    \centering
    \includegraphics[width=1.0\linewidth]{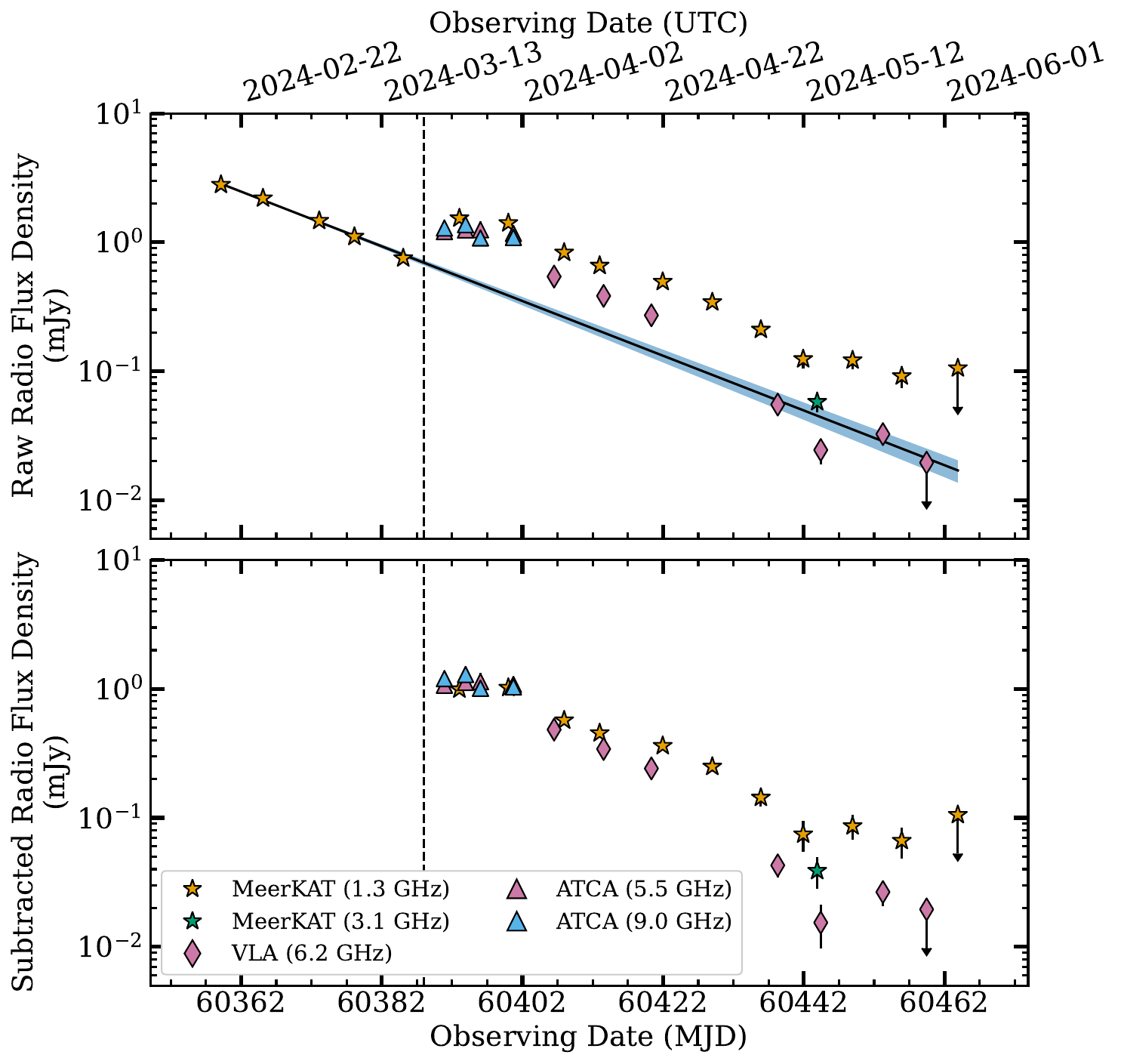}
    \caption{The decaying hard state: measured radio flux densities (\textit{top panel}); radio flux densities after subtraction of the exponentially decaying soft state component (\textit{bottom panel}). The solid black line and shaded blue area correspond to the exponential fit (to the 1.28\,GHz observations) and the $68\%$ confidence interval, respectively. The vertical dashed line marks the soft$\rightarrow$hard state transition \citep{Podgorny2024,Russel2024}.}
    \label{fig:LC_EXP}
\end{figure}

Figure~\ref{fig:LC_EXP} displays the exponential fit and the resulting residuals interpreted as compact jet emission. We address the potential for incomplete subtraction and its implications in the main text.

\section{Data Table}
Table~\ref{tab:LRLX_table} summarises the final radio and X-ray measurements presented in Figure~\ref{fig:lrlx}, assuming a distance of $D{=}2.6$\,kpc. The X-ray luminosities are interpolated to match the epochs of the radio observations. The table also includes the corresponding spectral indices and observing frequencies used to scale the radio luminosities to 5\,GHz.

\newpage
\def\arraystretch{1.5}
\begin{table*}
    \centering
    \begin{tabular}{llllll}
        \Xhline{3\arrayrulewidth}
        \multicolumn{1}{c}{Date\,(UTC)} & \multicolumn{1}{c}{Date\,(MJD)} & \multicolumn{1}{c}{$L_X$\,[1-10\,keV]\,(\ergs)} & \multicolumn{1}{c}{$L_R$\,[5\,GHz]\,(\ergs)} & \multicolumn{1}{c}{$\alpha$} & \multicolumn{1}{c}{$\nu_\text{obs}$\,(GHz)} \\ 
        \Xhline{3\arrayrulewidth}
2023-08-29T23:49:29 & 60185.993 & $(2.08_{-0.16}^{+0.15}){\times}10^{38}$ & $(4.8\,{\pm}\,0.5){\times}10^{30}$ & $-0.042\,{\pm}\,0.001$ & 5.20\\ 
2023-09-03T03:59:36 & 60190.166 & $(2.05_{-0.17}^{+0.16}){\times}10^{38}$ & $(4.3\,{\pm}\,0.4){\times}10^{30}$ & $+0.046\,{\pm}\,0.001$ & 4.75\\ 
2023-09-04T15:55:43 & 60191.664 & $(1.86_{-0.15}^{+0.15}){\times}10^{38}$ & $(3.9\,{\pm}\,0.4){\times}10^{30}$ & $+0.046\,{\pm}\,0.001$ & 1.28\\ 
2023-09-06T15:14:06 & 60193.635 & $(1.92_{-0.15}^{+0.15}){\times}10^{38}$ & $(4.1\,{\pm}\,0.4){\times}10^{30}$ & $+0.042\,{\pm}\,0.001$ & 1.28\\ 
2023-09-08T15:05:28 & 60195.629 & $(1.79_{-0.14}^{+0.14}){\times}10^{38}$ & $(3.7\,{\pm}\,0.4){\times}10^{30}$ & $+0.042\,{\pm}\,0.001$ & 1.28\\ 
2023-09-09T03:29:39 & 60196.146 & $(1.71_{-0.16}^{+0.16}){\times}10^{38}$ & $(3.8\,{\pm}\,0.4){\times}10^{30}$ & $+0.042\,{\pm}\,0.001$ & 4.75\\ 
2023-09-09T07:06:31 & 60196.296 & $(1.67_{-0.13}^{+0.12}){\times}10^{38}$ & $(4.0\,{\pm}\,0.4){\times}10^{30}$ & $-0.018\,{\pm}\,0.014$ & 5.50\\ 
2023-09-14T02:59:42 & 60201.125 & $(1.58_{-0.15}^{+0.15}){\times}10^{38}$ & $(3.8\,{\pm}\,0.4){\times}10^{30}$ & $+0.075\,{\pm}\,0.001$ & 4.75\\ 
2023-09-15T23:47:11 & 60202.991 & $(1.66_{-0.13}^{+0.13}){\times}10^{38}$ & $(3.9\,{\pm}\,0.4){\times}10^{30}$ & $+0.062\,{\pm}\,0.072$ & 5.20\\ 
2023-09-16T15:30:40 & 60203.646 & $(1.68_{-0.16}^{+0.16}){\times}10^{38}$ & $(4.0\,{\pm}\,0.4){\times}10^{30}$ & $+0.108\,{\pm}\,0.001$ & 1.28\\ 
2023-09-17T02:29:36 & 60204.104 & $(1.61_{-0.12}^{+0.12}){\times}10^{38}$ & $(3.9\,{\pm}\,0.4){\times}10^{30}$ & $+0.108\,{\pm}\,0.001$ & 5.75\\ 
\hline
2024-03-21T21:45:25 & 60390.907 & $(2.22_{-0.16}^{+0.15}){\times}10^{36}$ & $(4.2\,{\pm}\,0.5){\times}10^{28}$ & $+0.2\,{\pm}\,0.3$ & 5.50\\ 
2024-03-24T00:52:42 & 60393.037 & $(2.01_{-0.15}^{+0.14}){\times}10^{36}$ & $(4.7\,{\pm}\,0.7){\times}10^{28}$ & $+0.12\,{\pm}\,0.07$ & 5.50\\ 
2024-03-24T21:54:30 & 60393.913 & $(1.95_{-0.14}^{+0.14}){\times}10^{36}$ & $(4.5\,{\pm}\,0.5){\times}10^{28}$ & $+0.12\,{\pm}\,0.07$ & 1.28\\ 
2024-03-27T00:45:35 & 60396.032 & $(1.80_{-0.14}^{+0.13}){\times}10^{36}$ & $(4.6\,{\pm}\,0.9){\times}10^{28}$ & $+0.02\,{\pm}\,0.08$ & 5.50\\ 
2024-03-30T23:48:29 & 60399.992 & $(1.55_{-0.13}^{+0.13}){\times}10^{36}$ & $(4.4\,{\pm}\,0.6){\times}10^{28}$ & $+0.04\,{\pm}\,0.07$ & 6.22\\ 
2024-03-31T17:09:30 & 60400.715 & $(1.51_{-0.13}^{+0.13}){\times}10^{36}$ & $(4.4\,{\pm}\,0.4){\times}10^{28}$ & $+0.04\,{\pm}\,0.07$ & 1.28\\ 
2024-04-06T12:01:02 & 60406.501 & $(1.14_{-0.09}^{+0.09}){\times}10^{36}$ & $(2.0\,{\pm}\,0.2){\times}10^{28}$ & $-0.17\,{\pm}\,0.09$ & 1.28\\ 
2024-04-07T22:16:08 & 60407.928 & $(1.05_{-0.10}^{+0.10}){\times}10^{36}$ & $(1.8\,{\pm}\,0.3){\times}10^{28}$ & $-0.17\,{\pm}\,0.09$ & 6.22\\ 
2024-04-12T23:27:14 & 60412.977 & $(8.6_{-0.8}^{+0.7}){\times}10^{35}$ & $(1.5\,{\pm}\,0.3){\times}10^{28}$ & $-0.17\,{\pm}\,0.1$ & 6.22\\ 
2024-04-13T12:54:07 & 60413.538 & $(8.4_{-0.8}^{+0.8}){\times}10^{35}$ & $(1.44\,{\pm}\,0.15){\times}10^{28}$ & $-0.17\,{\pm}\,0.1$ & 1.28\\ 
2024-04-20T07:56:32 & 60420.331 & $(5.9_{-0.6}^{+0.6}){\times}10^{35}$ & $(1.04\,{\pm}\,0.12){\times}10^{28}$ & $-0.28\,{\pm}\,0.09$ & 1.28\\ 
2024-04-21T22:31:32 & 60421.939 & $(5.5_{-0.4}^{+0.4}){\times}10^{35}$ & $(9.99\,{\pm}\,0.18){\times}10^{27}$ & $-0.28\,{\pm}\,0.09$ & 1.28\\ 
2024-04-28T23:41:07 & 60428.987 & $(1.10_{-0.10}^{+0.10}){\times}10^{35}$ & $(6.89\,{\pm}\,0.13){\times}10^{27}$ & $-0.28\,{\pm}\,0.09$ & 6.22\\ 
2024-05-05T21:26:16 & 60435.893 & $(6.2_{-0.7}^{+0.7}){\times}10^{33}$ & $(2.6\,{\pm}\,0.8){\times}10^{27}$ & $-0.60\,{\pm}\,0.17$ & 1.28\\ 
2024-05-08T07:00:37 & 60438.292 & $(2.1_{-0.3}^{+0.3}){\times}10^{33}$ & $(2.0\,{\pm}\,0.4){\times}10^{27}$ & $-0.60\,{\pm}\,0.17$ & 3.06\\ 
2024-05-11T21:47:40 & 60441.908 & $(7_{-3}^{+5}){\times}10^{32}$ & $(8.0\,{\pm}\,0.4){\times}10^{26}$ & $-1.0\,{\pm}\,0.3$ & 6.22\\ 
2024-05-13T21:32:47 & 60443.898 & $(6_{-3}^{+4}){\times}10^{32}$ & $(9.8\,{\pm}\,0.3){\times}10^{26}$ & $-1.0\,{\pm}\,0.3$ & 1.28\\ 
2024-05-14T09:44:32 & 60444.406 & $(6_{-2}^{+3}){\times}10^{32}$ & $(7.7\,{\pm}\,0.3){\times}10^{26}$ & $-1.0\,{\pm}\,0.3$ & 6.22\\ 
2024-05-18T22:20:57 & 60448.931 & $(5_{-1.3}^{+1.6}){\times}10^{32}$ & $(1.5\,{\pm}\,0.5){\times}10^{27}$ & $-0.6\,{\pm}\,0.2$ & 1.28\\ 
2024-05-23T05:51:22 & 60453.244 & $(3.0_{-0.8}^{+1.1}){\times}10^{32}$ & $(1.2\,{\pm}\,0.3){\times}10^{27}$ & $-0.6\,{\pm}\,0.2$ & 6.22\\ 
2024-05-25T22:10:15 & 60455.924 & $(1.5_{-0.7}^{+1.6}){\times}10^{32}$ & $(1.1\,{\pm}\,0.4){\times}10^{27}$ & $-0.6\,{\pm}\,0.2$ & 1.28\\ 
2024-05-29T10:54:34 & 60459.455 & $(9.0_{-4}^{+8}){\times}10^{31}$ & $(3\,{\pm}\,2){\times}10^{26}$ & $-0.6\,{\pm}\,0.2$ & 6.00\\
        \Xhline{3\arrayrulewidth}
    \end{tabular}
    \caption{Summary of radio and X-ray properties presented in Figure~\ref{fig:lrlx} (assuming $D{=}2.6\,$kpc). Columns list the observation date (UTC), Modified Julian Date (MJD), X-ray luminosity in the 1--10 keV band ($L_X$), interpolated onto the radio observing epochs, radio luminosity at 5 GHz ($L_R$), radio spectral index ($\alpha$), and the observing radio frequency ($\nu_{\rm obs}$). X-ray luminosities are based on \textit{Swift}-XRT data, and radio values are extrapolated to 5 GHz assuming the listed spectral indices. The top section corresponds to the rise hard state; the bottom, the decay hard state. Uncertainties represent 1$\sigma$ errors.}
    \label{tab:LRLX_table}
\end{table*}


\bsp	
\label{lastpage}
\end{document}